\documentstyle[12pt]{article}

\begin{document}

\begin{flushright}
hep-th/0012238 \\
December 2000
\end{flushright}

\begin{centering}
\bigskip
{\leftskip=2in \rightskip=2in
{\large \bf Testable scenario for Relativity with minimum-length}}\\
\bigskip
\bigskip
\bigskip
{\bf Giovanni AMELINO-CAMELIA}\\
\bigskip
Dipartimento di Fisica, Universit\'{a} ``La Sapienza", P.le Moro 2,
I-00185 Roma, Italy\\ 
\end{centering}

\vspace{1cm}
\begin{center}
{\bf ABSTRACT}
\end{center}

{\leftskip=0.6in \rightskip=0.6in

I propose a general class of space-times whose structure is governed
by observer-independent scales of both velocity 
($c$) and length (Planck length),
and I observe that these space-times can naturally host
a modification of FitzGerald-Lorentz contraction such that 
lengths which in 
their inertial rest frame are bigger than a ``minimum length" are
also bigger than the minimum length in all other inertial frames. 
With an analysis in leading order in the minimum length,
I show that this is the case in a specific illustrative example
of postulates for Relativity with velocity and length 
observer-independent scales.
}

\newpage
\baselineskip 12pt plus .5pt minus .5pt
\pagenumbering{arabic}
\pagestyle{plain} 

The recent anniversary~\cite{jackqm100,zeilqm100,gacqm100} 
of Planck's introduction of his (reduced) 
constant $\hbar$ ($\hbar \simeq 10^{-34}J s$)
renders somewhat more disappointing the fact
that we have not yet established which role (if any)
should be played in the structure of space-time
by one of the implications of the
existence of $\hbar$ which appeared to be most significant to Planck:
the possibility to define the length scale now called 
Planck length $L_p$ by combining $\hbar$ with the 
gravitational constant $G$ and the speed-of-light constant $c$
($L_p \equiv \sqrt{\hbar G/c^3} \sim 1.6 {\cdot} 10^{-35}m$).
The fact that $L_p$ is proportional to both $\hbar$ and $G$
appears to invite one to speculate that it might play a role
in the microscopic (possibly quantum) 
structure of space-time, and in fact 
many ``quantum-gravity" theories 
(theories attempting to unify general relativity 
and quantum mechanics~\cite{gacqm100,rovhisto}), 
have either assumed
or stumbled upon this possibility; however, a fundamental role
for $L_p$ in the structure of space-time appears to be conceptually
troublesome for one of the cornerstones
of Einstein's Special Relativity:
FitzGerald-Lorentz length contraction.
(It is noteworthy that length contraction had already been
proposed by FitzGerald~\cite{fitzy}
and Lorentz~\cite{lorenpap} well before Planck's introduction
of his length scale.)
The Relativity Principle demands that physical laws should
be the same in all inertial frames, including the law
that would attribute to the Planck length a fundamental role
in the structure of space-time, whereas,
according to FitzGerald-Lorentz length contraction, 
different inertial observers would attribute different
values to the same physical length.
The idea that the Planck length should play 
a truly fundamental\footnote{While it is clear 
that a truly fundamental
role for the Planck length in space-time structure is inconsistent
with the combination of the Relativity Principle and ordinary
Fitgerald-Lorentz
length contraction, it is of course 
possible~\cite{grbgac,gampul,dsrd14,emnnew}
that the Planck length be associated with some sort of background
in a way that is consistent with 
both the Relativity Principle and Fitgerald-Lorentz
length contraction. 
This would be analogous to the well-known 
special-relativistic description of the motion of an electron
in a background electromagnetic field. 
This physical context is described by different 
observers in a way that is consistent
with the Relativity Principle, but only when these
observers take into account
the fact that the background electromagnetic field
also takes different values in different inertial frames.
The electric and the magnetic components of the background
field are not observer-independent, but 
their combination affects the motion of the electron
in a way that is of course consistent with the Relativity Principle.
The Planck length could play a similar role in fundamental physics,
{\it i.e.} it could reflect the properties of a background,
but then the presence of such a background
would allow to single out a ``preferred" class of inertial frames
for the description of the short-distance structure of space-time.
In the present study I show that in addition
to this scenario, which introduces the Planck
length together with a preferred class of inertial frames, 
it is also possible to follow another scenario
for the introduction of the Planck length: 
this second option does not predict
preferred inertial observers  but does require 
a short-distance deformation of FitzGerald-Lorentz contraction.}
role in the 
structure of space-time appears to be in conflict with the
combined implications of
the Relativity Principle and Fitgerald-Lorentz length contraction.

This conclusion does not depend on the specific role played by the
Planck length in space-time structure.
Let me clarify this by considering two popular ideas and showing
that the issue emerges in both.
A first popular example, which is encountered in many quantum-gravity
approaches~\cite{mead,padma,da} 
(including\footnote{It is of course not very significant
for the point here being made that actually in string theory
the minimum length might not be given exactly by the Planck
length (it could be a few orders of magnitude 
bigger~\cite{vene,grossme,amaven} or 
smaller~\cite{kab} than $L_p$).} 
string theory~\cite{vene,grossme,amaven,kab}),
is the one of the Planck length  
playing the role of ``minimum length", 
setting a limit on the localization of events. 
In this case the possibility to single out 
a preferred class of inertial frames for the
description of space-time emerges as a result of the fact
that an event localized with $L_p$ accuracy in one inertial frame,
would be, according to FitzGerald-Lorentz contraction,
localized with subPlanckian accuracy in some other inertial frames.
A second example is provided by quantum-gravity 
approaches (see, {\it e.g.}, Refs.~\cite{grbgac,gampul})
predicting new-physics
effects that would be strong
for particles of wavelength of the order of the Planck length
but would be weak for particles of larger wavelengths,
such as the ones associated with deformed dispersion relations
of the type\footnote{Note that 
from this point onward I use
conventions with $\hbar = 1$.} 
$E^2 = c^2 p^2 {\pm} L_p^{n} c^{2-n} E^{n} p^2$.
Assuming the special-relativistic rules of transformation
of energy and momentum,
these dispersion relations would allow to select a preferred
class of inertial frames.

In the present study I show 
that it is possible to formulate the Relativity postulates 
in a way that does not lead to inconsistencies in the case of
space-times whose short-distance structure is governed
by observer-independent scales of velocity and length,
and that this new type of relativistic theories allows the introduction
of a ``minimum length" and/or a length-scale deformation 
of the dispersion relation, without giving rise
to a preferred class of inertial frames for the
description of space-time structure.
The emerging picture provides a rather intuitive revision
of FitzGerald-Lorentz contractions: boosts are essentially
undeformed (Lorentz boosts) when acting on large lengths,
but the contraction becomes ``softer" when 
boosts act on short lengths.
The scenario is also attractive in light of the fact that
it makes predictions that are testable with planned experiments
such as the GLAST gamma-ray space telescope~\cite{glast},
and it appears even plausible that certain outstanding 
experimental puzzles~\cite{nsblazars} 
in astrophysics, which have already been
tentatively interpreted~\cite{kifu,aus,ita,sato,gactp2}
as possibly representing a manifestation of
a new fundamental length scale, could eventually be understood
in terms of Relativity with two 
observer-independent scales.

The first step of my analysis is an acknowledgement
of the central role that observer-independent scales (or absence thereof)
already played in Galilean Relativity and Einstein's Special Relativity.
This will prove useful for my task of introducing an
observer-independent length scale.
The Relativity Principle demands that ``the laws of physics are
the same in all inertial frames" and clearly 
the implications of this principle
for geometry and kinematics depend very strongly on whether the
fundamental structure of space-time hosts 
fundamental scales of velocity and/or length. In fact, the introduction
of a fundamental scale is itself a physical law, and therefore the
Relativity Principle allows the introduction of such fundamental
scales only if the rules that relate the observations performed
by different inertial observers are structured in such a way that
all inertial observers can agree on the value and physical interpretation
of the fundamental scales. 
The Galileo/Newton rules of transformation between 
inertial observers can be easily obtained by combining the 
Relativity Principle with the assumption 
that there are no observer-independent
scales for velocity or length.
For example, without an observer-independent velocity scale,
there is no plausible alternative to the simple
Galilean law $v'=v_0+v$ of composition of velocities

Special Relativity describes the implications
of the Relativity Principle for the case in which there is
an observer-independent velocity scale. 
Einstein's second postulate can be naturally divided in two
parts: the introduction of an observer-independent
velocity scale $c$ ($c \simeq 3 {\cdot} 10^8 m/s$)
and the proposal of a physical interpretation of $c$ as the
speed of light. This second postulate, when combined with the 
Relativity Principle 
(which is the first postulate of Special Relativity)
and with the additional assumption
that there is no observer-independent length scale
leads straightforwardly to the now familiar Lorentz transformations,
with their associated familiar formulation 
of FitzGerald-Lorentz contraction.
The assumption that there is no observer-independent length scale
plays a key role already in the way in which the second postulate
was stated. Experimental data available when Special Relativity was
formulated, such as the ones of the Michelson-Morley experiments,
only concerned light of very long wavelengths (extremely long
in comparison with the length scale $L_p$ introduced by Planck
a few years earlier) and therefore the second postulate could
have accordingly attributed to $c$ the physical role of
speed of long-wavelength light (the infinite-wavelength limit
of the speed of light); however, the implicit assumption of
absence of an observer-independent length scale allowed
to extrapolate from Michelson-Morley data a property
for light of all wavelengths. 
In fact, it is not possible to assign
a wavelength dependence to the speed of light
without introducing a ``preferred" class of inertial frames 
or an observer-independent length scale.

All the revolutionary elements of Special Relativity
(in comparison with the Relativity of Galileo and Newton)
are easily understood as direct consequences of the
introduction of an observer-independent velocity scale
This is particularly clear for the deformed law
of composition of 
velocities, $v'=(v_0+v)/(1+v_0v/c^2)$,
and the demise of absolute time (which is untenable
when an observer-independent velocity
scale governs the exchange of information between clocks).

Within the perspective here being adopted it is clear
that the Planck-length problem I am concerned with can be
described as the task of showing that the Relativity
Principle can coexist with the following postulate
\begin{itemize}
\item{(L.1):} The laws of physics 
involve a fundamental velocity scale $c$
and a fundamental length scale $L_p$.
\end{itemize}
The addition of an observer-independent length scale
does not require
major revisions of the physical interpretation of $c$, but, 
because of the mentioned connection 
between wavelength independence and absence of an observer-independent
length scale, I shall not authomatically assume that it is
legitimate to extrapolate from our long-wavelength data:
\begin{itemize}
\item{(L.1b):} The value of the fundamental velocity scale $c$
can be measured by each inertial observer as 
the $\lambda/L_p \rightarrow \infty$ limit of the speed of light
of wavelength $\lambda$.
\end{itemize}

While for $c$ we can at least rely on long-wavelength data,
we basically have no experimental information on the role (if any)
of $L_p$ in space-time structure.
I can only use the intuition that is emerging from quantum-gravity
approaches. I shall focus on the two 
mentioned popular ideas: $L_p$ could play the 
role of ``minimum length" 
or the role of a reference scale for wavelengths, characterizing
deformed dispersion relations. Among the results reported
in the present study the one which appears to be most compelling 
to this author
is the fact that the requirement of consistency with 
the Relativity Principle
(and absence of a preferred frame for the description
of the short-distance structure of space-time)
can provide a connection between these 
otherwise unrelated intuitions: in some scenarios in which
the Planck length is a reference scale of wavelengths characterizing
deformed dispersion relations one can derive from consistency
with the Relativity Principle that the Planck length
is also the ``minimum length". 
Motivated by the objective
of describing this connection
and by the desire to provide an analysis that is relevant
for planned~\cite{glast}
experimental studies of the possibility of dispersion relations of 
type $E^2 = c^2 p^2 {\pm} L_p c E p^2$,
I consider here the following example of possible 
physical interpretation of the Planck length:
\begin{itemize}
\item{(L.1c):} Any inertial observer can establish the value 
of $L_p$ by determining the dispersion relation for photons, 
which takes the form $E^2 = c^2 p^2 + f(E,p;L_p)$, where 
the function $f$ has leading $L_p$ dependence given 
by: $f(E,p;L_p) \simeq L_p c E p^2$.
\end{itemize}

The first task for establishing the logical consistency
of this illustrative example 
of new Relativity postulates is the one of showing
that there is a satisfactory deformation of Lorentz
transformations such that the dispersion 
relation $E^2 \simeq c^2 p^2 + L_p c E p^2$
holds in all inertial frames for fixed (observer-independent)
value of $L_p$.
It is actually quite easy (although it involves somewhat tedious
mathematics) to construct these deformed transformations.
I shall give a detailed technical description of the
derivation and of the general properties of the
deformed transformation rules elsewhere~\cite{dsrd14}.
For the analysis of the most significant 
physical implications of the new postulates
it is sufficient to note here the transformation rules
for boosts of photon momentum along the direction of motion.
Specifically, 
let us consider a photon which, for a given inertial observer,
is moving along the positive direction of the $z$ axis with
momentum $p_0$ (and, of course, as imposed by the new dispersion
relation, has 
energy $E_0 \simeq p_0 + L_p c p_0^2/2$).
The new relativity postulates imply~\cite{dsrd14} that 
for another inertial observer, which the first observer sees moving
along the same $z$ axis, the photon has momentum $p$
related to $p_0$ by
\begin{equation}
p = p_0 e^{-\xi} 
+ L_p p_0^2 e^{- \xi} 
- L_p p_0^2 e^{- 2 \xi} ~,
\label{finitep}
\end{equation}
where $\xi$ is the familiar rapidity parameter of boosts.
As manifest in (\ref{finitep}), ordinary Lorentz
boosts are of course
obtained as the $L_p \rightarrow 0$ limit of the new boosts.
The comparison between (\ref{finitep})
and its $L_p \rightarrow 0$ limit also allows 
us to gain some insight on the type of deformation of 
FitzGerald-Lorentz contraction that characterizes the new postulates,
and the associated emergence of a ``minimum length".
As long as $p_0 < 1/L_p$ (wavelength $\lambda_0>L_p$)
and $e^{-\xi} \ll 1/(L_p p_0)$ the relation between $p$ and $p_0$
is well described by ordinary Lorentz transformations.
Within the analysis in leading order in $L_p$ here reported
it is not legitimate to consider the case $e^{-\xi} > 1/(L_p p_0)$
(which would require 
an exact all-order analysis of the implications of the
function $f(E,p;L_p)$ introduced in the postulates), 
but we can look at the behaviour
of the transformation rules when $e^{-\xi}$ is smaller but not
much smaller than $1/(L_p p_0)$.
While the transformation rules are basically unmodified 
when $e^{-\xi} \ll 1/(L_p p_0)$, as $e^{-\xi}$
approaches from below the value $1/(L_p p_0)$ 
the transformation rules are more and
more severely modified: for large boosts,
the ones that would lead to nearly Planckian wavelengths
in the ordinary special-relativistic case, the magnitude of
the wavelength contraction is sizably reduced.
For example, for $e^{-\xi} \simeq 1/(3 L_p p_0)$ one
would ordinarily predict $p \simeq 1/(3 L_p)$ while the new
transformation rules predict the softer momentum $p \simeq 2/(9 L_p)$.
This suggests that there should exist an
exact all-order form of $f(E,p;L_p)$
(extending the present $f(E,p;L_p) \simeq L_p c E p^2$ leading-order
analysis) such that when one inertial observer assigns to the photon
momentum smaller than $1/L_p$ (wavelength greater than $L_p$)
all other inertial observers also find momentum smaller than $1/L_p$.

In order to gain some more direct
intuition on the new type of length contraction
which can emerge in theories with
observer-independent scales of velocity and length, 
it is useful to analyse a gedanken length-measurement procedure.
A key point for these analyses is the fact that the 
dispersion relation $E^2 \simeq c^2 p^2 + L_p c E p^2$ 
corresponds~\cite{grbgac,schaef,billetal}
to the deformed speed-of-light law
\begin{equation}
v_\gamma(p) = c \, (1 + L_p |p|/2)
~.
\label{eqvelocity}
\end{equation}
The wavelength dependence of this speed-of-light law
plays a key role in the emergence of a minimum length
in measurement analysis.
I show this in a simple context. Let us consider two observers
each with its own (space-) ship moving in the same space direction, 
the $z$-axis, with different
velocities ({\it i.e.} with some relative velocity),
and let us mark ``A" and ``B" 
two $z$-axis points on one of the ships (the rest frame).
The procedure of measurement of the distance $AB$ is structured as 
a time-of-flight measurement:
an ideal mirror is placed at B
and the distance is measured
as the half of the time needed by a first photon wave packet,
centered at momentum $p_0$, sent from A toward B to be back at A
(after reflection by the mirror).
Timing is provided by a digital light-clock: another mirror is placed
in a point ``C" of the rest frame/ship, with the same $z$-axis coordinate
of A at some distance $AC$, and a second identical\footnote{The two
wave packets are taken here to be identical only for simplicity.
Nothing prevents one from considering a wave packet for the
light-clock with (central) momentum $p_0^*$ ($p_0^* \ne p_0$).
This more general case will be analysed elsewhere~\cite{inprep}
emphasizing the fact that the 
possibility $p_0^* \ne p_0$ plays a role in the
difference between high precision
measurements of large distances and 
low-precision measurements of short distances.}
wave packet,
again centered at $p_0$, is bounced back and forth between A and C.
The rest-frame observer will therefore measure $AB$ 
as $AB'=v_\gamma(p_0) {\cdot} N {\cdot} \tau_0/2$, where $N$ is the number of 
ticks done by
the digital light-clock during the A$\rightarrow$B$\rightarrow$A
journey of the first wave packet
and $\tau_0$ is the time interval corresponding to each
tick of the light-clock ($\tau_0 =2 \, AC/v_\gamma(p_0)$).
The observer on the second (space-) ship, moving with velocity $V$
with respect to the rest frame, will instead attribute to $AB$
the value
\begin{equation}
AB''= {v_\gamma(p)^2 - V^2 \over v_\gamma(p)} N {\tau \over 2}
~,
\label{dsfitzloa}
\end{equation}
\begin{equation}
AB''= {[v_\gamma(p)^2 - V^2] v_\gamma(p_0) \over 
v_\gamma(p) \sqrt{v_\gamma(p')^2 - V^2}} 
N {\tau_0 \over 2} 
= {v_\gamma(p)^2 - V^2 \over 
v_\gamma(p) \sqrt{v_\gamma(p')^2 - V^2}}  AB'
~,
\label{dsfitzlob}
\end{equation}
where $p$ is related to $p_0$ through (\ref{finitep}),
while $p'$ is related to $p_0$ through the corresponding
formula~\cite{inprep} for boosts in a direction orthogonal to
the one of motion of the photon.
The derivation of (\ref{dsfitzloa})-(\ref{dsfitzlob}) 
is completely analogous
to the derivation of the familiar special-relativistic formulas
in the analysis of the same measurement
procedure assuming wavelength-independence of the speed of light,
but the new
ingredient of the wavelength dependence of the speed of light
has important physics implications.
In particular, (\ref{dsfitzlob}) reflects
time dilatation in the new relativistic theory ($\tau$ is the  
time interval which the second observer, moving with respect to
the rest frame, attributes to each
tick of the light-clock).

The implications of (\ref{dsfitzlob}) for length contraction
are in general quite complicated, but they are easily analysed in both
the small-$V$ and the large-$V$ limits
(examined here of course in leading order in $L_p$).
For small $V$ and small momentum (large wavelength) of the probes
Eq.~(\ref{dsfitzlob}) reproduces ordinary FitzGerald-Lorentz contraction.
For large $V$ Eq.~(\ref{dsfitzlob}) 
predicts that $AB''$ receives two most important
contributions: the familiar FitzGerald-Lorentz 
term ($AB' {\cdot} \sqrt{c^2-V^2}$) and a new term 
which is positive and of order $L_p |p| AB'/ \sqrt{c^2-V^2}$.
As $V$ increases the ordinary FitzGerald-Lorentz contribution
to $AB''$ decreases as usual, but the new correction term
increases. Imposing $|p| > |\delta p| > 1/AB''$ (the probe wavelength
must of course be shorter than the distance being measured)
one arrives at the 
result $AB'' > \sqrt{c^2-V^2}AB' + L_p AB'/(AB'' \sqrt{c^2-V^2})$,
which clearly is such that $AB'' > L_p$ for all values of $V$.
Again I must remind the reader that I am
here reporting an analysis in leading order in $L_p$, and therefore
the results cannot be trusted when $V$ is large enough that the
correction term is actually bigger
than the 0-th order contribution to $AB''$,
but we can trust the indications of this analysis as long
as the correction is smaller than the 0-th order term, and 
in that regime one finds that FitsGerald-Lorentz contraction
is being significantly softened in the region corresponding 
to nearly Planckian contraction.
This result clearly supports the hypothesis that there should
exist a consistent all-order form of $f(E,p;L_p)$ such that when one 
inertial observer assigns to a length value greater than $L_p$ 
all other inertial observers also find that length to be greater
than $L_p$.
Such a form of $f(E,p;L_p)$ would provide a relativistic
theory with observer-independent scales $c$ and $L_p$ in which $L_p$
has the intuitive role of ``minimum length" described above.

For the future development of the general type of Relativity theories
here proposed, it is important
to understand whether this result is a general
prediction of Relativity with oberver-independent $c$ and $L_p$
or it requires the specific formulation of the postulates here explored.
Even assuming that, as here proposed, space-time structure is
characterized by observer-independent $c$ and $L_p$, and that
it is consistent with the Relativity Principle (and that it
does not involve some associated new background which singles out
a preferred class of inertial observers for the description
of space-time structure), one could contemplate
a wide spectrum of possible roles for $L_p$, some involving
different 
leading-order forms of the dispersion relation,
some not even describable as deformations of the dispersion
relation. Let me postpone to future studies this latter possibility,
and consider here the former possibility:
other scenarios for the leading-order form of the dispersion relation.

One first case to be considered is the one in which the 
leading-order form of the deformation
is just the same as the one here examined but with opposite
sign: $f(E,p;L_p) \simeq - L_p c E p^2$. In that case all formulas
here obtained would, of course, still be valid upon changing all
the signs of the coefficients of $L_p$, and it is easy
to see that a minimum length would not arise. (Changing the relevant
signs one finds that the predicted contraction is even stronger
than the FitzGerald-Lorentz one, and boosts reach subPlanckian
lengths even more quickly than in Special Relativity.)
So it appears that a minimum length requires that the wavelength
dependence of the speed of light is such that bigger values
of the wavelength lead to higher velocities. (It is perhaps worth
noting that this author started, long ago, these studies with
an unjustified but strongly felt intuition that the opposite
situation should be favoured, an intuition which was changed
by these results on minimum length.)

A second case which should be considered for illustrative purposes
in the one in which $f(E,p;L_p)$ is such that the leading order
is only quadratic in the 
Planck length, {\it e.g.} $f(E,p;L_p) \simeq L_p^2 E^2 p^2$
or $f(E,p;L_p) \simeq L_p^2 E^4/c^2$. In these cases (since the signs
have been chosen positive) one does easily find the same qualitative
results here obtained, with the only difference that the effects
are weaker. In particular, these scenarios do lead to the emergence of
a minimum length, but the region of good validity (validity to very
good approximation) of ordinary FitzGerald-Lorentz contraction
is somewhat extended with respect to the case on which I focused.
The qualitative behaviour is however, exactly the same: 
the contraction is basically underformed for small velocities
of the second observer with respect to the rest frame, but it
gets softened for large velocities in a way that forbids
access to subPlanckian lengths.
It appears therefore that the overall sign of the 
leading-order deformation of the dispersion relation is fixed
by the requirement of a minimum length, while the power of $L_p$
appearing in the leading-order term cannot be fixed by
such a requirement.

My closing remarks concern the important subject of 
the phenomenological
implications of this proposal of new Relativity postulates.
Since the Planck length is so tiny, and the effects here
considered are inevitably suppresed by the ratio of the
Planck length versus the wavelength of the photon or
versus the length being measured,
one might be tempted to assume that none of these effects
could be tested in the near future.
However, this is not true, at least not
for equation (\ref{eqvelocity}).
As emphasized in Refs.~\cite{grbgac,schaef,billetal},
experiments that will be done in a few years ({\it e.g.}
by GLAST~\cite{glast}) are expected to achieve sensitivity levels
sufficient for tests\footnote{As mentioned,
the scenario for wavelength dependence considered 
in Refs.~\cite{grbgac,schaef,billetal} 
would not make room for $L_p$ in the Relativity postulates,
but the type of wavelength
dependence is the one here considered (with the important
physical/observable difference 
that there would be preferred inertial observers
for the description of the dispersion relation).}
of the possibility
of $L_p/\lambda$ wavelength dependence of the speed of light.
The fact that linear effects can be studied in the near future is 
of course very significant for the Relativity proposal I am making,
but a high-priority issue for the development of this research programme
appears to be the one of finding experimental strategies 
for tests of quadratic effects
(if the leading
order of the deformation is quadratic in $L_p$ the effect
would be too small for presently-known
strategies of experimental studies of the velocity law).

In this respect, while waiting for these tests of 
the $L_p$-linear scenario of velocity-law deformation,
it appears most urgent to develop a general analysis of threshold
energies in the new Relativity framework.
As mentioned, certain outstanding 
experimental puzzles~\cite{nsblazars} 
in astrophysics appear~\cite{kifu,aus,ita,sato,gactp2}
to invite one to introduce a deformation of special-relativistic
threshold energies involving a new length scale.
In fact, the relevant astrophysics data appear to be
inconsistent with the special-relativistic
evaluation of the (threshold) energies
required by certain processes.
Various authors~\cite{kifu,aus,ita,sato,gactp2}
have observed that the paradox can be solved by
introducing a new length scale in way that would allow
the identification of a preferred class of inertial frames.
It would be exciting to discover that the general 
type of relativistic
frameworks here proposed (in which a new length scale is
introduced in a way that does not involve preferred inertial frames)
also provides a solution of the threshold paradoxes.
However, this requires a careful analysis of
massive particles in the new framework, which is not easily
done in full generality. For example, in the case on which I primarily
focused here, where photons 
satisfy $E^2 \simeq c^2 p^2 + L_p c E p^2$,
massive particles should consistently satisfy a dispersion
relation of the type $E^2 \simeq c^4 m^2 + c^2 p^2 + F(E,p;m;L_p)$,
with $F$ some function whose combined limit small-$L_p$,$m \! = \! 0$
is given by $L_p c E p^2$
(and, of course, since $c^2 m$ is the rest energy of the 
particle, $F(0,E;m;{\tilde L}_p)=0$). 
Elsewhere~\cite{dsrd14,inprep} I shall show
that assuming that $F$ is $m$-independent one obtains a consistent
relativistic theory, but I shall also show that within that
assumption one does not obtain an explanation of the
threshold paradoxes. I shall however observe that certain types
of $m$ dependence of $F$ would provide a solution of the paradoxes, 
but these reacher structures of $F$ are
not easily analyzed with respect to the consistency of
the type of relativistic theory here proposed.
Work on general criteria for the consistency of the function $F$
appear to be strongly motivated by the present analysis,
since they might lead to the interpretation of the
threshold paradoxes observed in astrophysics 
as the first manifestation of an
observer-independent minimum length in Nature.

\baselineskip 12pt plus .5pt minus .5pt

\vfil


\begin{thebibliography}{99}

\bibitem{jackqm100} D.~Kleppner and R.~Jackiw, Science 289 (2000) 893.

\bibitem{zeilqm100} A.~Zeilinger, Nature 408 (2000) 639.

\bibitem{gacqm100} G.~Amelino-Camelia, gr-qc/0012049, Nature 408 (2000) 661.

\bibitem{rovhisto} C.~Rovelli,
gr-qc/0006061.

\bibitem{fitzy} G.F.~FitzGerald,
{\it The ether and the Earth's atmosphere}
(Science, 1889) 

\bibitem{lorenpap} H.A.~Lorentz,
{\it Versuch einer Theorie der elektrischen und optischen
Erscheinungen von bewegten Koerpern} 
(Leiden, 1895). 

\bibitem{grbgac} G. Amelino-Camelia, J. Ellis, N.E. Mavromatos, 
D.V. Nanopoulos and S. Sarkar, 
astro-ph/9712103,
Nature {393} (1998) 763.

\bibitem{gampul} R.~Gambini and J.~Pullin,
Phys.~Rev.~D59 (1999) 124021.

\bibitem{dsrd14} G. Amelino-Camelia, gr-qc/0012051.

\bibitem{emnnew} J. Ellis, N.E. Mavromatos and D.V. Nanopoulos,
hep-th/0012216.

\bibitem{mead} C.A.~Mead,
Phys.~Rev.~135 (1964) B849.

\bibitem{padma} T.~Padmanabhan,
Class.~Quantum Grav.~4 (1987) L107.

\bibitem{da} D.~V.~Ahluwalia,
Phys.~Lett.~B339 (1994) 301.

\bibitem{vene} G.~Veneziano, Europhys.~Lett.~2 (1986) 199.

\bibitem{grossme} D.J.~Gross and P.F.~Mende,
Nucl.~Phys.~B303 (1988) 407.

\bibitem{amaven} D.~Amati, M.~Ciafaloni and G.~Veneziano, 
Phys.~Lett.~B216 (1989) 41.

\bibitem{kab} D.~Kabat and P.~Pouliot, Phys.~Rev.~Lett.~77 (1996) 1004;
M.R.~Douglas, D.~Kabat, P.~Pouliot and S.H.~Shenker, 
Nucl.~Phys.~B485 (1997) 85.

\bibitem{glast} J.P.~Norris, J.T.~Bonnell, G.F.~Marani, 
J.D.~Scargle, 
astro-ph/9912136;
A.~de Angelis,
astro-ph/0009271.

\bibitem{nsblazars} H.~Muir, {\it Blazars}, 
New Sci.~(23 September 2000) 32-35.
     
\bibitem{kifu} T.~Kifune,
astro-ph/9904164, Astrophys.~J.~Lett.~518 (1999) L21.

\bibitem{aus} R.J.~Protheroe and H.~Meyer,
Phys.~Lett.~B493 (2000) 1.

\bibitem{ita} 
R.~Aloisio, P.~Blasi, P.L.~Ghia and A.F.~Grillo,
Phys.~Rev.~D62 (2000) 053010.


\bibitem{sato} H.~Sato,
astro-ph/0005218.

\bibitem{gactp2} G. Amelino-Camelia and T. Piran, astro-ph/0008107.

\bibitem{schaef} B.E.~Schaefer,
Phys.~Rev~Lett.~82 (1999) 4964.

\bibitem{billetal} S.D. Biller {\it et al},
Phys.~Rev.~Lett.~83 (1999) 2108.

\bibitem{inprep} G. Amelino-Camelia, in preparation.

\end{thebibliography}
\end{document}